\documentclass[preprint,pra,superscriptaddress,floatfix]{revtex4-1}

\usepackage{graphicx}
\usepackage{amssymb}
\usepackage{amsmath}
\usepackage{latexsym}

\begin{document}

\title{Non-locally entangled microwave and micromechanical squeezed cats: a phase transition-based protocol}

\author{A. A. Gangat}
\thanks{a.gangat@physics.uq.edu.au}
\affiliation{ARC Centre for Engineered Quantum Systems, School of Mathematics and Physics, The University of Queensland, St Lucia, QLD 4072, Australia}
\author{I. P. McCulloch}
\affiliation{ARC Centre for Engineered Quantum Systems, School of Mathematics and Physics, The University of Queensland, St Lucia, QLD 4072, Australia}
\author{G. J. Milburn}
\affiliation{ARC Centre for Engineered Quantum Systems, School of Mathematics and Physics, The University of Queensland, St Lucia, QLD 4072, Australia}
\begin{abstract}
{\bf Electromechanical systems currently offer a path to engineering quantum states of microwave and micromechanical modes that are of both fundamental and applied interest.  Particularly desirable, but not yet observed, are mechanical states that exhibit entanglement, wherein non-classical correlations exist between distinct modes; squeezing, wherein the quantum uncertainty of an observable quantity is reduced below the standard quantum limit; and Schr\"odinger cats, wherein a single mode is cast in a quantum superposition of macroscopically distinct classical states.  Also, while most investigations of electromechanical systems have focussed on single- or few-body scenarios, the many-body regime remains virtually unexplored.  In such a regime quantum phase transitions naturally present themselves as a resource for quantum state generation, thereby providing a route toward entangling a large number of electromechanical systems in highly non-classical states.  Here we show how to use existing superconducting circuit technology to implement a (quasi) quantum phase transition in an array of electromechanical systems such that entanglement, squeezing, and Schr\"odinger cats become simultaneously observable across multiple microwave and micromechanical oscillators.}
\end{abstract}

\maketitle

Observing the quantum behaviour of macroscopic mechanical objects has received serious consideration in the literature since the turn of the century \cite{Cleland, Bose, Blencowe1, Carr, Armour, Marshall, Blencowe2, Schwab}, and in 2010 O'Connell et al. \cite{O'Connell} demonstrated for the first time a non-classical state of a macroscopic mechanical object by placing the dilational mode of a micromechanical resonator in a quantum superposition of the ground state and the first excited state.  To date, however, mechanical entanglement, mechanical squeezing, and superpositions of "macroscopically distinct" \cite{Leggett} mechanical states (mechanical Schr\"{o}dinger cats) remain unobserved.  Non-local mechanical entanglement, first considered by Mancini et al. \cite{Mancini}, is of fundamental interest as it would further elucidate the nature of the quantum-classical boundary and may provide the opportunity for teleporting the quantum state of the centre of mass of a macroscopic object \cite{TianCarr}.  From an applied perspective, the force-sensing capability of mechanical oscillators may be enhanced by entanglement \cite{metrology}, and the expected long coherence times of mechanical oscillators at low temperatures may make them useful for quantum information processing (QIP).  Squeezed mechanical states \cite{Blencowe1}, where the quantum noise in one quadrature of the mechanical oscillator is reduced below the standard quantum limit, in addition to being of interest in exploring the quantuam-classical boundary can also be useful for metrology \cite{BAE}.  Finally, mechanical Schr\"{o}dinger cats are relevant to tests of various models of environmentally induced decoherence \cite{Zurek} in mechanical resonators \cite{Schlosshauer, Remus} and tests of gravitational and non-gravitational wavefunction collapse theories (see \cite{Bassi} for a review or \cite{Pepper} for a summary).

ECSs (see \cite{Sanders_review} for a review) are able to combine Schr\"{o}dinger cats and non-local entanglement in a single multipartite state such that, "The individual coherent states retain the desirable quasiclassical properties of coherent states but nonlocal features arise due to the entanglement" \cite{SandersECS}.  It is therefore highly desirable to create ECSs of mechanical resonators with squeezed coherent states as this would simultaneously encapsulate mechanical entanglement, squeezing, and Schr\"{o}dinger cats.  Bose and Agarwal proposed a scheme \cite{BoseAgarwal} to create an ECS with two nanomechanical resonators (NRs) coupled to a cooper pair box (CPB), and Tian and Zoller \cite{TianZoller} considered ECS creation in the situation of two NRs coupled to a trapped ion.  However, both schemes are not readily scalable to larger numbers of resonators, do not involve squeezed states, and remain experimentally underdeveloped.
 
In the photonic domain there has been much theoretical interest in ECS generation and use at optical frequencies \cite{Sanders_review}, and the first experimental demonstration of ECS generation was published in 2009 by A. Ourjoumtsev et al. \cite{opticalECS} for the case of two distant optical modes.  Generating ECSs in superconducting microwave circuits, however, has received little attention.  Superconducting circuits are an important platform for QIP and scale much more easily to large numbers of cavities than do optical setups.  This is particularly significant for W-ECSs (ECSs of the form $|\alpha, 0, 0,...0,0\rangle+|0,\alpha, 0,...0,0\rangle+...+|0, 0, 0,...0,\alpha\rangle$) as they can be understood as superpositions of discrete variable W-states, for which the bipartite and global entanglement decay under phase damping and the global entanglement decay under amplitude damping have been shown to be independent of the number of qubits \cite{W-state decay}.  To our knowledge there have been no proposals to generate W-ECSs in microwave circuits to date.

Here we present the first scheme to deterministically prepare large amplitude ($|\alpha|^2\gg1$) squeezed multipartite ECSs in an array of coupled superconducting microwave coplanar waveguides (CPWs) and to subsequently transfer them to an array of uncoupled micromechanical oscillators.  Mechanical mode entanglement is thereby achieved by simple state swapping as in the proposal of \cite{Braunstein}.  We employ the technique of entanglement generation via a quantum phase transition of a many-body Hamiltonian \cite{BECcat,Lee,Plenio}, which naturally accommodates a large number of oscillators and is particularly well-suited to the platform of superconducting circuits \cite{Houck}.  Our scheme achieves squeezing of the coherent states through Kerr nonlinearities in the CPWs.  The resulting ECSs are able to take the following forms:

\begin{align}
|\Psi_{W-sECS} \rangle &= \frac{1}{\sqrt{R}} \sum_{j=1}^R |\alpha_{sq}\rangle_j \prod_{r\neq j}|0\rangle_r, \\
~\nonumber \\
|\Psi_{W-sESCS} \rangle &= \frac{1}{\sqrt{R}} \sum_{j=1}^R \frac{1}{\sqrt{2}}\Big(e^{-i\pi/4}|i\alpha_{sq}\rangle_j \\
&~~~~~~+e^{i\pi/4}|-i\alpha_{sq}\rangle_j \Big) \prod_{r\neq j}|0\rangle_r,  \nonumber \\
|\Psi_{GHZ-sECS} \rangle &= \frac{1}{\sqrt{2}}e^{-i\pi/4}\prod_{j}\Big|i\frac{\alpha_{sq}}{\sqrt{M}}\Big\rangle_j \nonumber \\
&~~~~~~+ \frac{1}{\sqrt{2}}e^{i\pi/4}\prod_{j}\Big|-i\frac{\alpha_{sq}}{\sqrt{M}}\Big\rangle_j \label{GHZ-sECS}
\end{align}

respectively the W-type squeezed ECS (W-sECS), W-type squeezed entangled Schrodinger cat state (W-sESCS), and GHZ-type squeezed ECS (GHZ-sECS), where $R$ is the total number of modes over which the ECS is distributed, M is the total number of CPWs, $|\alpha_{sq}\rangle$ denotes a quadrature squeezed coherent state of amplitude $\alpha$, the subscripts for the GHZ-sECS are CPW indices, and $j$ and $r$ are mode indices such that each mode may be a single fundamental mode of a CPW or the fundamental normal mode of a chain of CPWs.  Our scheme is therefore the first to be able to distribute ECSs amongst the fundamental normal modes of distinct oscillator arrays \cite{MECSnote}, thereby producing non-locally superposed microwave/micromechanical coherent state superfluids.  Also, the ECSs, in addition to being either purely microwave or purely micromechanical, may be distributed amongst a combination of microwave and micromechanical oscillators.  We design our scheme around existing technology, as we describe below, and find that parameters accessible to present-day devices can yield a purely mechanical version of the ECSs across ten oscillators with a multipartite superposition coherence of greater than sixty percent for an initial coherent state of amplitude $|\alpha|^2\approx10$.  We also perform a numerical verification of our scheme under amplitude and phase damping with the master equation formalism using realistic parameters.

\section*{System and model}

\begin{figure*} [htp]
\begin{center}
\includegraphics[scale=0.3, angle=0]{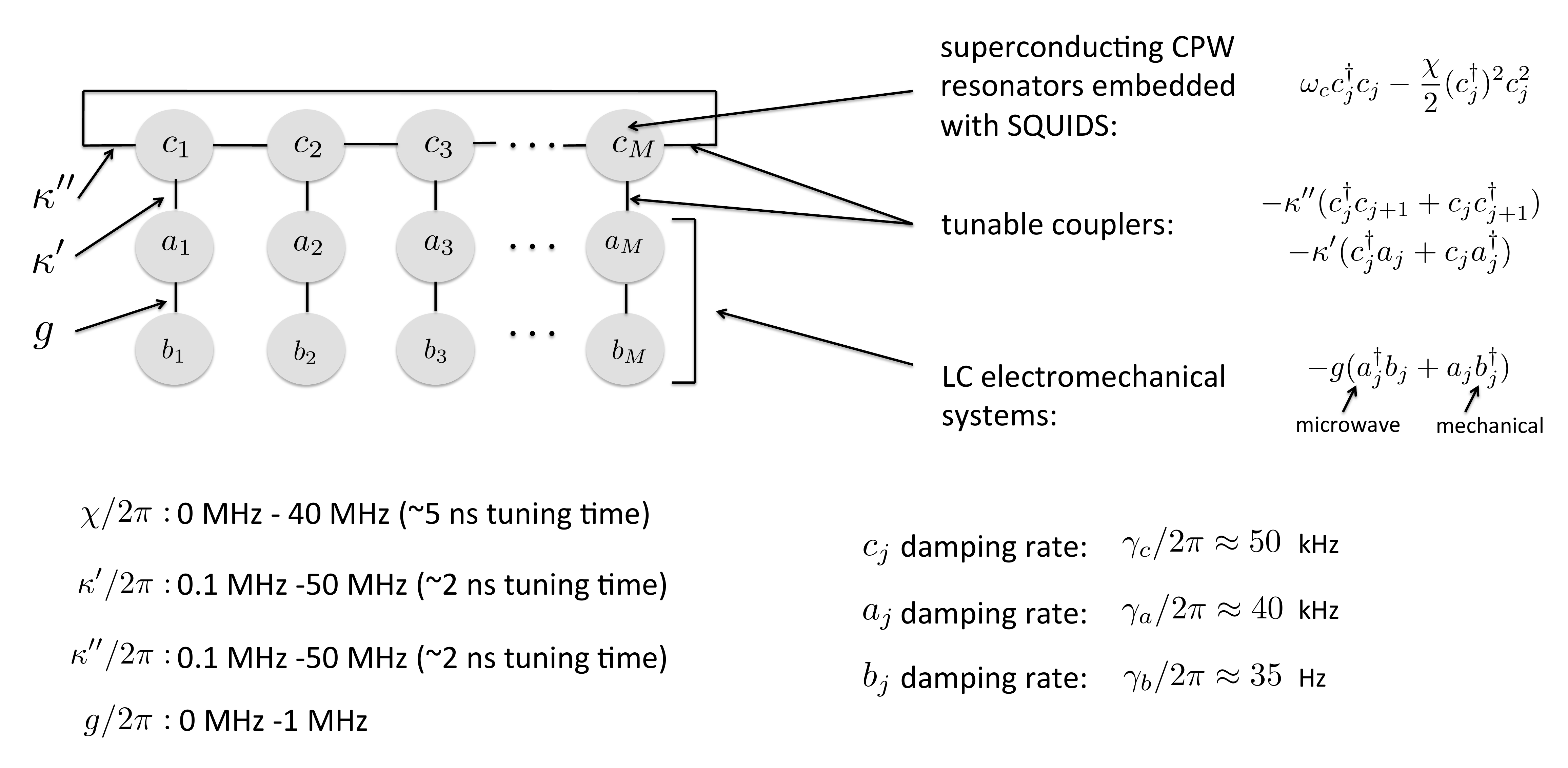}
\end{center}
\caption{Proposed system and parameters.  Grey circles denote microwave CPW ($c_j$), microwave lumped LC ($a_j$), and micromechanical ($b_j$) oscillator modes, whose parameters are taken from refs. \cite{Bourassa} and \cite{Teufel1}.  Solid black lines denote tunable inter-mode coupling.  The CPWs are embedded with transmon SQUIDS to give a negative Kerr nonlinearity to the $c_j$ modes (ref. \cite{Bourassa}).  The transmon has been experimentally shown to have a $T_1>1.5~\mu$s and a minimum $T_{\phi}$ of $\geq35~\mu$s \cite{transmon1}.  The minimum value of $T_{\phi}$ occurs at the flux sweet spot, but considering an asymmetric SQUID as in \cite{Bourassa} allows for $T_{\phi}\gtrsim1$ ms at the flux values where $\chi$ is minimum and maximum \cite{Bourassa2}.  Each CPW is tunably coupled to an LC mode of an electromechanical system of the type demonstrated in refs. \cite{Teufel1, Teufel2} and  nearest neighbour CPWs by the tuneable coupler of refs. \cite{coupler,coupler2}.  Each electromechanical system has its own capability for tuneable coupling between its LC and mechanical modes \cite{Teufel1}. (These parameters entail $\langle c^{\dagger}c \rangle \leq 10$.  A larger limit for $\langle c^{\dagger}c \rangle$ can be achieved with larger $\omega_a$ and maximum $\omega_c$; see Supplementary Information.)} 
\label{system} 
\end{figure*}

We consider a one-dimensional chain of sites with periodic boundary conditions (Fig. \ref{system}).  Each site consists of a superconducting microwave frequency CPW resonator, and the microwave LC electromechanical system demonstrated by Teufel et al. in \cite{Teufel1, Teufel2}.  A transmon qubit SQUID is embedded in the CPW resonator as experimentally demonstrated first by J. M. Fink et al. \cite{transmon_expts}.  The theoretical analysis in \cite{Bourassa} shows how this can induce a Kerr nonlinearity in the fundamental microwave mode $c$ of the CPW: $H_c=\hbar \omega_c c^{\dagger}c - \hbar \frac{\chi}{2}c^{\dagger}c(c^{\dagger} c - 1)$; $\omega_c$ and $\chi$, respectively the frequency and Kerr constant of the mode, are dependent on the flux through the SQUID loop, with $\omega_c$ achieving a minimum for the flux value that maximizes $\chi$ and vice versa.  (Higher order terms of the nonlinearity can be neglected if $\chi\ll\omega_{c,0}/2\langle c^{\dagger}c\rangle$, where $\omega_{c,0}$ is the fundamental mode frequency when $\chi=0$; see Supplementary Information).  The electromechanical system can achieve the quantum-coherent strong-coupling interaction Hamiltonian $H_{int}=-\hbar g(a^{\dagger} b + a b^{\dagger})$, with $g$ as the interaction strength, which allows state swapping between the fundamental mechanical mode $b$ and LC microwave mode $a$ as recently demonstrated \cite{Palomaki}.  The fundamental CPW mode $c$ is coupled to the LC mode $a$ via the tunable coupler demonstrated in \cite{coupler,coupler2}, and $\omega_c=\omega_a$ when $\chi=0$ so that state swapping can occur between $a$ and $c$ also.  The chain is formed by linking the fundamental modes $c_j$ of nearest neighbour CPWs, where $j$ is the site index, also with the tunable coupler.  All of the relevant dynamical time scales for our proposed system are much shorter than the dissipation time scales of the relevant degrees of freedom (see Fig. \ref{system}). 

The model Hamiltonian for such a system, when the modes $a_j$ are all driven with equal and sufficient strength at frequency $\omega_D$ near their red mechanical sidebands in the "resolved sideband" limit \cite{Milburn2012}, can be expressed after appropriate transformations and approximations as
\begin{equation}
\begin{split}
\frac{H}{\hbar} = \sum_{j=1}^M \delta_j a^{\dagger}_j a_{j} &+ (\delta_j-\Delta_{j}) c^{\dagger}_j c_{j} - \frac{\chi_{j}}{2}c_j^{\dagger}c_j(c_{j}^{\dagger} c_{j} - 1) \\
&- \kappa'(c_{j}^{\dagger} a_{j} + c_{j} a_{j}^{\dagger}) - \kappa''_j(c_{j}^{\dagger} c_{j+1} +  c_{j}c_{j+1}^{\dagger}) \\
& - g(a_{j}^{\dagger} b_{j} + a_{j} b_{j}^{\dagger}),
\end{split}
\label{Ham}
\end{equation}
where $j$ is the site index for a total of $M$ sites, $c_{M+1}=c_1$ (periodic boundary conditions), $\delta_j=\omega_{a,j} - \omega_{b,j} - \omega_{D,j}$ is the detuning of the drives from the red mechanical sidebands, $\Delta_{j}=\omega_{a,j} - \omega_{c,j}$, $g=g_0\sqrt{n_D}$, $g_0$ is the bare cavity optomechanical coupling, and $n_D$ is the number of drive photons in the resonators of the modes $a_j$.  The Hamiltonian preserves the total number of quanta in the system.  Because of this and the fact that the dissipation time scales are much longer than the relevant dynamical time scales of the unitary evolution, we may consider the qualitative physics of the system in an isolated and number-conserving picture.  

With uniform parameters ($\delta_j=\delta$, $\Delta_j=\Delta$, $\chi_j=\chi$, $\kappa''_j=\kappa''$), equation (\ref{Ham}) is essentially the Hamiltonian for the one dimensional ABH with two additional degrees of freedom ($a_j$ and $b_j$) linked to each lattice site.  For $\kappa'\ll\kappa''$, these additional degrees of freedom are effectively decoupled from the lattice, and we may make a unitary transformation to recover the ABH that is theoretically studied in Refs. \cite{BECcat,attractiveBH1,attractiveBH2,attractiveBH3,attractiveBH4,attractiveBH5,attractiveBH6,attractiveBH7,attractiveBH8,attractiveBH9}:
\begin{equation}
\frac{H}{\hbar} = \sum_{j=1}^M -\frac{\chi}{2}c_j^{\dagger}c_j(c_{j}^{\dagger} c_{j} - 1) - \kappa''(c_{j}^{\dagger} c_{j+1} +  c_{j}c_{j+1}^{\dagger}).
\label{ABH_Ham}
\end{equation}
In this way our proposed system is similar to that of \cite{Lieb} and provides a more robust alternative platform to BECs in optical lattices \cite{attractiveBH1, attractiveBH2}, coupled atom-cavity systems \cite{Plenio}, and trapped ions \cite{attractiveBH6} for the experimental realization of the ABH.  We also note that the combination of quartic and quadratic terms in equation (\ref{ABH_Ham}), given sufficient time-dependent control over the $\chi_j$ and $\kappa_j$ and supplemented by coherent displacements of the fields, should enable one to sue this system as a universal Bosonic simulator based on a result of Braunstein and Lloyd \cite{BL}.  In the Supplementary Information we demonstrate this for the case of bipartite W-ECS creation.

In the ABH the ground state changes qualitatively as a function of the parameter $\tau=\frac{\kappa''}{\chi (N-1)}$, where $N$ is the total number of quanta and we assume $N>1$.  Any particular realization of the ABH can be characterized by the constants $\tau_1$ and $\tau_2$ ($\tau_2\geq\tau_1$) such that for $\tau>\tau_2$  the ground state is a superfluid, for $\tau_1>\tau$ the ground state is a Schr\"{o}dinger cat-like state, and the intermediate regime $\tau_2>\tau>\tau_1$ is transitional in nature \cite{attractiveBH2}.  $\tau_1\approx0.25$ is largely independent of lattice size \cite{attractiveBH2,attractiveBH4}, while $\tau_2\approx[2Msin^2(\pi/M)]^{-1}$ for $M>5$ and $0.25<\tau_2\lesssim0.3$ for $3\leq M \leq5$ \cite{attractiveBH3}.  $\tau_2=\tau_1$ for $M=2$ \cite{attractiveBH4}.  In this paper we are concerned with accessing the Schr\"{o}dinger cat regime ($\tau<\tau_1$), where, for decreasing values of $\tau$, the ground state approaches the W-state $|\Psi_{N}\rangle = \frac{1}{\sqrt{M}} \sum_{j=1}^M|N\rangle_j\prod_{r\neq j}|0\rangle_r$, the on-site number fluctuations ($\langle c_{j}^{\dagger}c_{j}^{\dagger}c_{j}c_{j} \rangle - \langle c_{j}^{\dagger}c_{j} \rangle^2$) approach $N^2\sqrt{M-1}/M$, and the single particle correlation between sites ($\langle c_j^{\dagger}c_{j+1} \rangle$) approaches zero \cite{attractiveBH1,attractiveBH2}.

In \cite{Plenio} Brandao et al. suggested using an atom-cavity realization of the ABH to create a polaritonic (photon-atom hybrid) W-state in multiple optical cavities via adiabatic transition from the superfluid regime to the Schr\"{o}dinger cat regime by tuning $\chi$ (we note that this is not strictly a quantum phase transition as the thermodynamic limit for the ABH is ill-defined \cite{attractiveBH4}).  This is possible also for the microwave quanta in our proposed system as the experimental parameter space that we specified above allows a minimum $\tau$ value of $0.0025/(N-1)$.  By extension, however, the linearity of quantum mechanics allows a number distribution in the form of a coherent state $|\alpha\rangle=e^{-|\alpha|^2/2}\sum_{n=0}^\infty \frac{\alpha^n}{\sqrt{n!}}|n\rangle$ (with amplitude $\alpha$ large enough such that $|\langle 1|\alpha\rangle|^2\lesssim0.01$) to become a site-localized superposition of generalized coherent states: $|\Psi_{\alpha_{gen}}\rangle = \frac{1}{\sqrt{M}} \sum_{j=1}^M |\alpha_{gen}\rangle_j\prod_{r\neq j}|0\rangle_r=\frac{1}{\sqrt{M}} \sum_{j=1}^M e^{-|\alpha|^2/2}\sum_{n=0}^\infty e^{i\phi_n}\frac{\alpha^n}{\sqrt{n!}}|n\rangle_j\prod_{r\neq j}|0\rangle_r$, where a number-dependent phase factor $e^{i\phi_n}$ appears for each number state component due to the number-dependent frequency induced by the Kerr medium (cyclic phase shearing of a coherent state in a single Kerr medium was demonstrated in the recent experiment of G. Kirchmair et al. \cite{Kerr_experiment} for the case of fixed $\chi$).  For a coherent state of amplitude $\alpha$ localized on a single site, we have the exact relation $\phi_n=n\int\frac{\chi(t)}{2}dt$.  For $M>1$ the situation is complicated by the hopping term in the Hamiltonian and the fact that each number component sees a different value of $\tau$ for the same value of $\chi$.  However, our numerical investigations show that the different number components periodically come within a small enough phase difference of each other after the overall phase transition such that the coherent state at each site is periodically revived as a squeezed coherent state $|\alpha_{sq}\rangle$ or squeezed Schr\"{o}dinger cat state $\frac{1}{\sqrt{2}}(e^{-i\pi/4}|i\alpha_{sq}\rangle+e^{i\pi/4}|-i\alpha_{sq}\rangle)$ to yield the W-sECS/W-sESCS/GHZ-sECS.

A more general form of the site-localized states $|\Psi_{\alpha_{gen}}\rangle$ is $|\Psi_{\alpha_{gen}}^{(block)}\rangle\\=\frac{1}{\sqrt{R}} \sum_{l=1}^R |\alpha_{gen}\rangle_l\prod_{r\neq j}|0\rangle_r$, where the lattice is divided into a total of $R$ adjacent blocks with an arbitrary number of sites in each (see Fig. (\ref{stateprep}) for an example), $l$ and $r$ are block indices, and $|\alpha_{gen}\rangle_l$ designates a generalized coherent state of amplitude $\alpha$ occupying the lowest energy normal mode of lattice block $l$: $|\alpha_{gen}\rangle_l=e^{-|\alpha|^2/2}\sum_{n=0}^\infty e^{i\phi_n}\frac{\alpha^n}{n!}(a_{l,k=0}^{\dagger})^n|0\rangle_l$, where $a_{l,k=0}^{\dagger}$ is the creation operator for the lowest energy normal mode of block $l$.  This cat state generalizes the W-ECSs studied in \cite{WECS1,WECS2,WECS3,WECS4}, which consist of coherent states confined to single local modes but entangled between multiple local modes, to W-ECS where any component of the superposition may be distributed over multiple spatially adjacent local modes.  Such states may be thought of as non-local superpositions of superfluids.  We next discuss the experimental protocol of how to prepare $|\Psi_{\alpha_{gen}}^{(block)}\rangle$ in our system and convert it into a purely mechanical state.

\section*{Preparation protocol}

\begin{figure*} [htp]
\begin{center}
\includegraphics[scale=0.47]{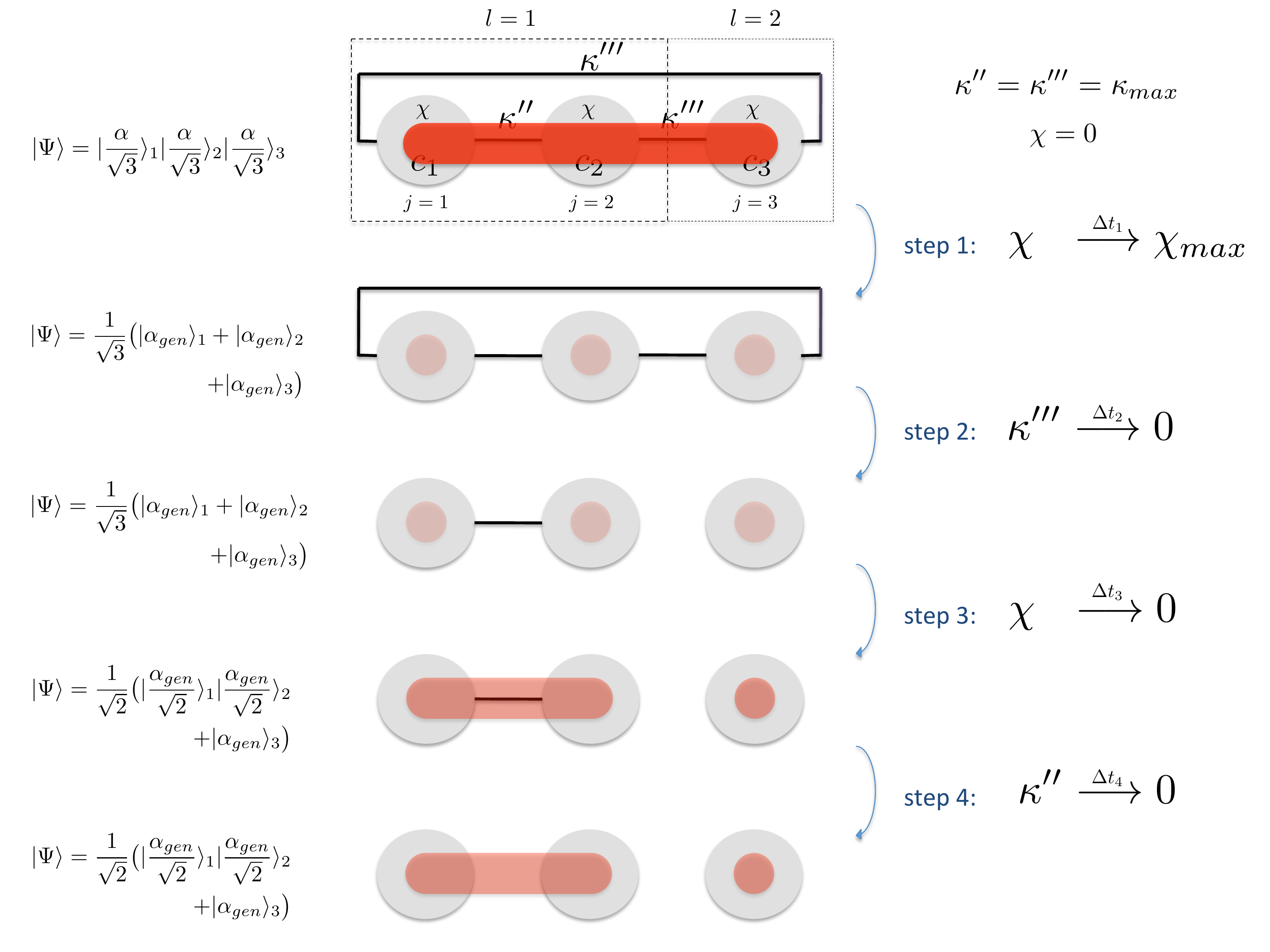}
\end{center}
\caption{Steps one, two, three, and four of the state preparation protocol illustrated for the case of a total of three sites ($M=3$) in two blocks ($R=2$).  Grey circles denote microwave CPW oscillator modes ($c_j$) with negative Kerr nonlinearities of strength $\chi$.  Solid black lines denote tunable intra-block ($\kappa''$) and inter-block ($\kappa'''$) couplings.  The coherent state indicated in red shading can be site-localized (circle) or distributed over adjacent sites (oblong).  Darker shading intensity reflects higher occupation probability.  An initial coherent state is created in the lowest energy normal mode of the lattice, then converted into a site-localized superposition $|\Psi_{\alpha_{gen}}\rangle$ by adiabatically increasing $\chi$ in step one.  Due to the presence of the Kerr medium, the initial coherent state $|\alpha\rangle$ acquires a time-dependent shearing of its Wigner function and is therefore labeled as a generalized coherent state $|\alpha_{gen}\rangle$.  In step two the blocks are decoupled by tuning $\kappa'''$ to zero.  In step three $\chi$ is adiabatically tuned back to zero to convert the state into the block-localized superposition $|\Psi_{\alpha_{gen}}^{(block)}\rangle$.  In step four the sites within each block are decoupled and $|\Psi_{\alpha_{gen}}^{(block)}\rangle$ is then ready to be transferred to the mechanical modes $b_j$ via state-swapping with modes $a_j$.  The phase shearing of $|\alpha_{gen}\rangle$ ceases at the end of step three when $\chi=0$, and with appropriate timing the final state of each block corresponds to a superposition of the vacuum state $|0\rangle$ and either a squeezed coherent state $|\alpha_{sq}\rangle$ or a squeezed Schr\"{o}dinger cat state $\frac{1}{\sqrt{2}}(e^{-i\pi/4}|i\alpha_{sq}\rangle_j+e^{i\pi/4}|-i\alpha_{sq}\rangle_j)$.}
\label{stateprep} 
\end{figure*}

The experimental protocol for preparing $|\Psi_{\alpha_{gen}}^{(block)}\rangle$ in the $c_j$ modes is illustrated in Fig. (\ref{stateprep}) for the case of $M=3$ and $R=2$.  The couplings $\kappa''$ that cross block boundaries are specially designated as $\kappa'''$.  At the beginning of the protocol, $\chi_j=0$, and the microwave modes $a_j$ and $c_j$ are all in their ground states at cryogenic temperatures and are uncoupled by setting $\kappa'=0.1$ MHz.  After setting $\kappa''_j=\kappa''_{max}=40$ MHz, a coherent drive is applied to the lowest energy normal mode of the $c_j$ lattice to create a coherent state of amplitude $\alpha$, after which the drive is turned off.  This results in a product coherent state of the local modes: $|\frac{\alpha}{\sqrt{M}}\rangle_1|\frac{\alpha}{\sqrt{M}}\rangle_2...|\frac{\alpha}{\sqrt{M}}\rangle_M$.  The $n=1$ component of the normal mode coherent state will not undergo the phase transition as $\tau$ is increased, so we require $\alpha$ sufficiently large such that $P_{1}=|\langle 1|\alpha\rangle|^2$ is very small.

The first step of the protocol adiabatically transforms the multimode coherent state of the $c_j$ modes into $|\Psi_{\alpha_{gen}}\rangle$ by increasing $\chi$.  From the discussion in the first section, we know that for increasing values of $\chi/\kappa''$ the ground state with $N=n$ is transformed at the critical point $\chi/\kappa''\approx4/(n-1)$ to a Schr\"{o}dinger cat-like state.  Considering the largest value of $\chi$ as $\chi_{max}/2\pi=40$ MHz, the bound on the smallest non-negligible Fock state component $|n_{min}\rangle$ of the coherent state must be $n_{min}\geq\frac{4\kappa_{max}''}{\chi_{max}}+1=5$.  If we impose the constraint $\sum_{n=0}^{4}|\langle n|\alpha\rangle|^2 \leq 0.03$, we find that $|\alpha|^2\geq10$ is sufficient to be able to neglect number state components with $n<5$.

Tuning $\tau$ through the transition (by changing $\chi$) to achieve $|\Psi_{\alpha_{gen}}\rangle$ involves contending with two dynamical processes of the system \cite{hybridmode}.  In the superfluid and intermediate regimes ($\tau>\tau_1$), the hopping terms of the Hamiltonian (equation (\ref{ABH_Ham})) proportional to $\kappa''$ allow the quanta to redistribute into a site-localized superposition as $\tau$ is tuned.  Adiabaticity with respect to $\kappa''$ must therefore be maintained while $\chi\leq\frac{4\kappa''_{max}}{n-1}$.  From Fig. 7 of \cite{attractiveBH2} we approximate the adiabatic condition for changing $\chi$ as $\frac{d\chi}{dt}\lesssim\frac{\kappa_{max}''^2}{10\tau_2 (n-1)}$, which implies $\Delta t_{1}\gtrsim0.040\tau_2~\mu$s for all $n\geq5$, where $\Delta t_{1}$ is the time taken to adiabatically tune $\chi$ within the regime $\tau>\tau_1$.  After the adiabatic tuning, $\chi$ can be non-adiabatically increased to $\chi_{max}$.  By increasing $\chi$ while $\kappa''$ is maximal, $\Delta t_{1}$ is minimized and therefore the effect of the transmon's minimal dephasing time in the vicinity of the flux "sweet spot" can be minimized.  On the other hand, as discussed in \cite{attractiveBH1}, because perfect degeneracy of the lattice sites is not experimentally feasible there will be non-uniformity of the lattice population induced in the site basis at a rate $\Delta\epsilon/\hbar$, where $\Delta\epsilon$ is the difference between on-site energies.  This is relevant in the intermediate and Schr\"{o}dinger cat regimes ($\tau_2>\tau$).  If the intermediate regime is traversed on a time scale comparable to or longer than $\hbar/\Delta\epsilon$, the system will have time to transition to the global ground state of complete localization in the lowest energy site.  We therefore require $\Delta t_{int}\ll\hbar/\Delta\epsilon$, where $\Delta t_{int}$ is the time it takes to traverse the intermediate regime, so that the spatially uniform distribution of the quanta from the superfluid regime is maintained and the system ends up in a superposition of localization on each site.  For the conservative estimate of $\Delta t_{int}\approx\Delta t_1$, this implies $\Delta\epsilon/\hbar\lesssim2\pi\times4/\tau_2$ MHz, which can be satisfied for moderate lattice sizes ($2\leq M\leq80$) as reference \cite{Underwood} demonstrates fabrication capability for GHz frequency CPW resonators with variations $\Delta\omega_c\approx2\pi\times1$ MHz, while contributions to $\Delta\epsilon$ from intersite variations in $\chi_j$ can be minimized through experimental calibration of the control signals to keep the $\chi_j$ nearly equal during step one of the state preparation protocol.  

Note that starting with an initial Fock state (with $N>1$) instead of an initial coherent state would yield a W-state.  Our proposed scheme can therefore be an alternative to the one experimentally demonstrated by Wang et al. in \cite{WangNOON}, but with the added advantage of N being constrained by the number of sites M rather than qubit $T_2$ times: if each site is tunably coupled to a qubit, M quanta can be simultaneously loaded into the array (one in each mode $c_j$ via the scheme demonstrated in \cite{Hofheinz}) while $\chi,\kappa''=0$, after which $\kappa''$ can be adiabatically tuned to $\kappa_{max}$ to prepare the fixed number superfluid state as the initial state of step one.

After the first step, the state of the lattice is approximately $|\Psi_{\alpha_{gen}}\rangle$ (assuming $|\alpha|^2\geq10$).  The third step of the protocol expands the site-localized $|\alpha_{gen}\rangle$ of $|\Psi_{\alpha_{gen}}\rangle$ to the block-localized $|\alpha_{gen}\rangle$ of $|\Psi_{\alpha_{gen}}^{(block)}\rangle$.  $\chi$ is tuned back to $0$ (adiabatically for $\tau>0.25$) so that the site-localized $|\alpha_{gen}\rangle$ become block-localized superfluid $|\alpha_{gen}\rangle$ and the state of the $c_j$ lattice is $|\Psi_{\alpha_{gen}}^{(block)}\rangle$.  From the discussion of the first step we know that this must be done on a timescale $\Delta t_3\gtrsim\Delta t_{1}$ to maintain adiabaticity.  Further, as discussed toward the end of the first section, if $\alpha$ is sufficiently large the timing may be chosen such that $|\alpha_{gen}\rangle\approx|\alpha_{sq}\rangle$ or $|\alpha_{gen}\rangle\approx\frac{1}{\sqrt{2}}(e^{-i\pi/4}|i\alpha_{sq}\rangle+e^{i\pi/4}|-i\alpha_{sq}\rangle)$.  If there is only one block for the whole lattice ($R=1$), the latter case results in the GHZ-sECS of equation (\ref{GHZ-sECS}).

In the fourth step, the intra-block couplings $\kappa''/2\pi$ are rapidly tuned ($\Delta t_4\sim2$ ns) to $0.1$ MHz so that all the $c_j$ modes are decoupled.  This may alter the population distribution somewhat within each block due to imperfect diabaticity, but the superfluid nature (inter-site phase coherence) of the blocks is preserved as there is no onsite interaction.

In the final step, state transfer from $c_j$ to $a_j$ ($\Delta t_{5a}\approx0.01~\mu$s) then $a_j$ to $b_j$ ($\Delta t_{5b}\approx0.1$ $\mu$s) is  done by pulsing $\kappa'$ and $g$, respectively.  Although the mechanical modes ($b_j$) at cryogenic temperatures will have significant thermal occupation, the experiment of \cite{Palomaki} demonstrated that the state swap effectively cools the mechanical mode ($b_j$) by transferring the thermal quanta to the LC mode ($a_j$) so that the state of the mechanical mode ($b_j$) immediately after the swap contains the desired state plus about one residual thermal quantum.  For blocks with more than one oscillator, there is phase coherence between the oscillators of the block and the state of the block may be considered in this respect as a micromechanical superfluid.  We note that the conversion from microwave to micromechanical can also be site-specific instead of lattice-wide, resulting in an entangled microwave-micromechanical state.  The multipartite superposition coherence after step 5b can be greater than sixty percent for $M\lesssim10$ (see Supplementary Figure 1).

\section*{Numerical simulation}

\begin{figure} [htp]
\begin{center}
\includegraphics[scale=0.5]{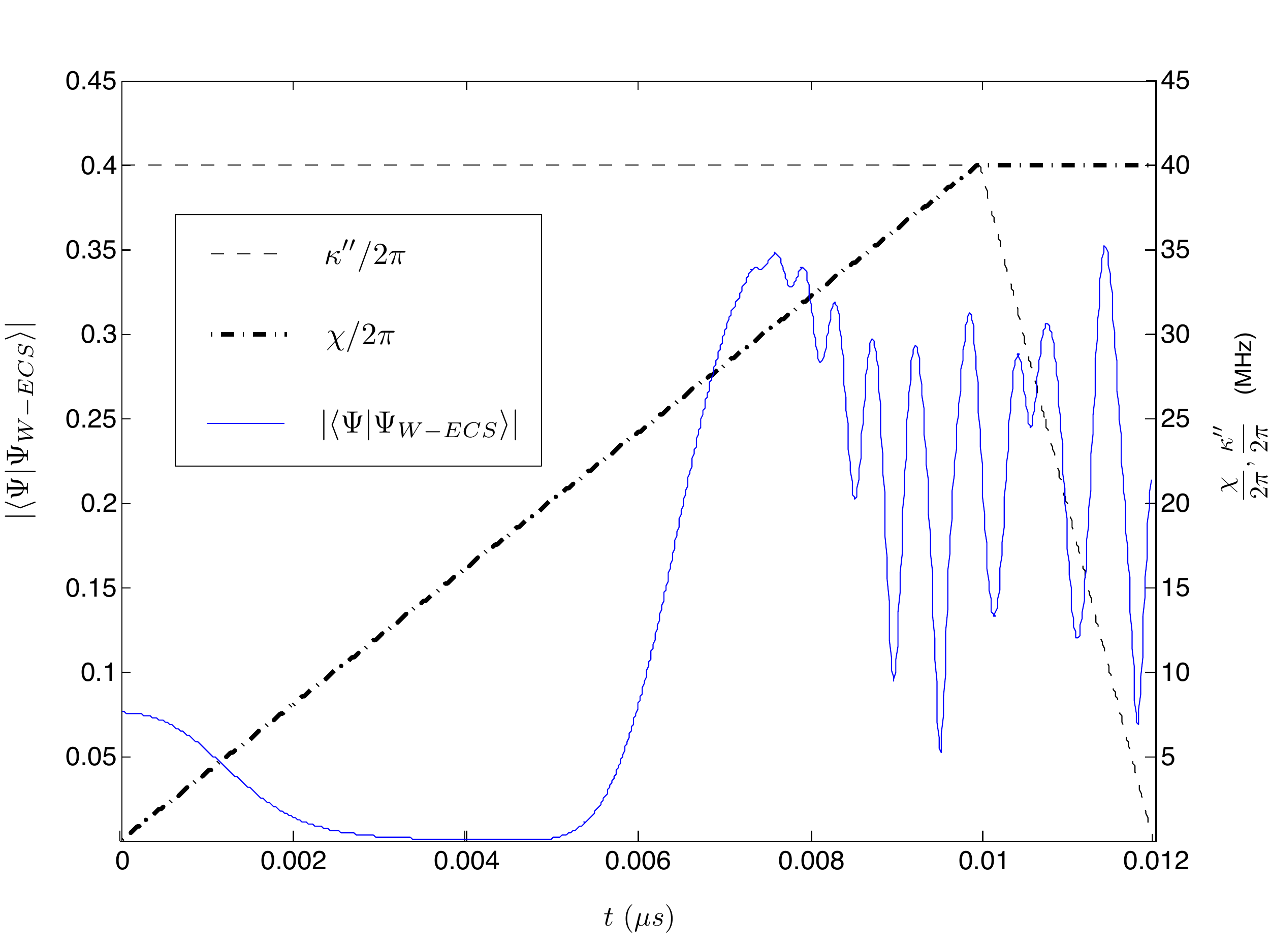}
\end{center}
\caption{Fidelity of the system state with a non-squeezed W-ECS over a simulation of steps one and two of the state preparation protocol for an initial coherent state of amplitude $|\alpha|^2=10$ in the lowest energy normal mode of a two site chain of CPWs.}
\label{fig_steps1and2} 
\end{figure}

We provide a proof of principle demonstration of our proposal through numerical integration of a master equation (see Methods) over steps one, two, and three of our state preparation protocol for two sites with realistic parameters as given above.  We model the $T_1$ and $T_{\phi}$ flux dependence linearly such that $T_1=3~\mu$s, $T_{\phi}=1$ s when $\chi=0$ and $T_1=1.5~\mu$s, $T_{\phi}=100~\mu$s when $\chi=\chi_{max}$.  A reference simulation without damping is also performed.  The initial state is a coherent state of amplitude $|\alpha|=\sqrt{10}$ in the lowest energy normal mode.  The timing of each step respects both the minimum tuning times of each device and the adiabaticity constraint.
Figure \ref{fig_steps1and2} shows the time-evolution of the fidelity ($|\langle\Psi|\Psi_{W-ECS}\rangle|$) through steps one and two, where $|\Psi\rangle$ is the damped system state and the reference state $|\Psi_{W-ECS}\rangle$ is a non-squeezed version of the W-sECS.  $\chi$ is tuned from zero to $\chi_{max}$ in 10 ns, after which $\kappa''$ is tuned to zero in 2 ns.  The oscillations of the fidelity are indicative of the cyclic phase shearing of the site-localized coherent state in phase space due to the nonlinearity in each site of the lattice.  The oscillations increase in frequency as $\chi$ is increased.

\begin{figure} [htp]
\begin{center}
\includegraphics[scale=0.5]{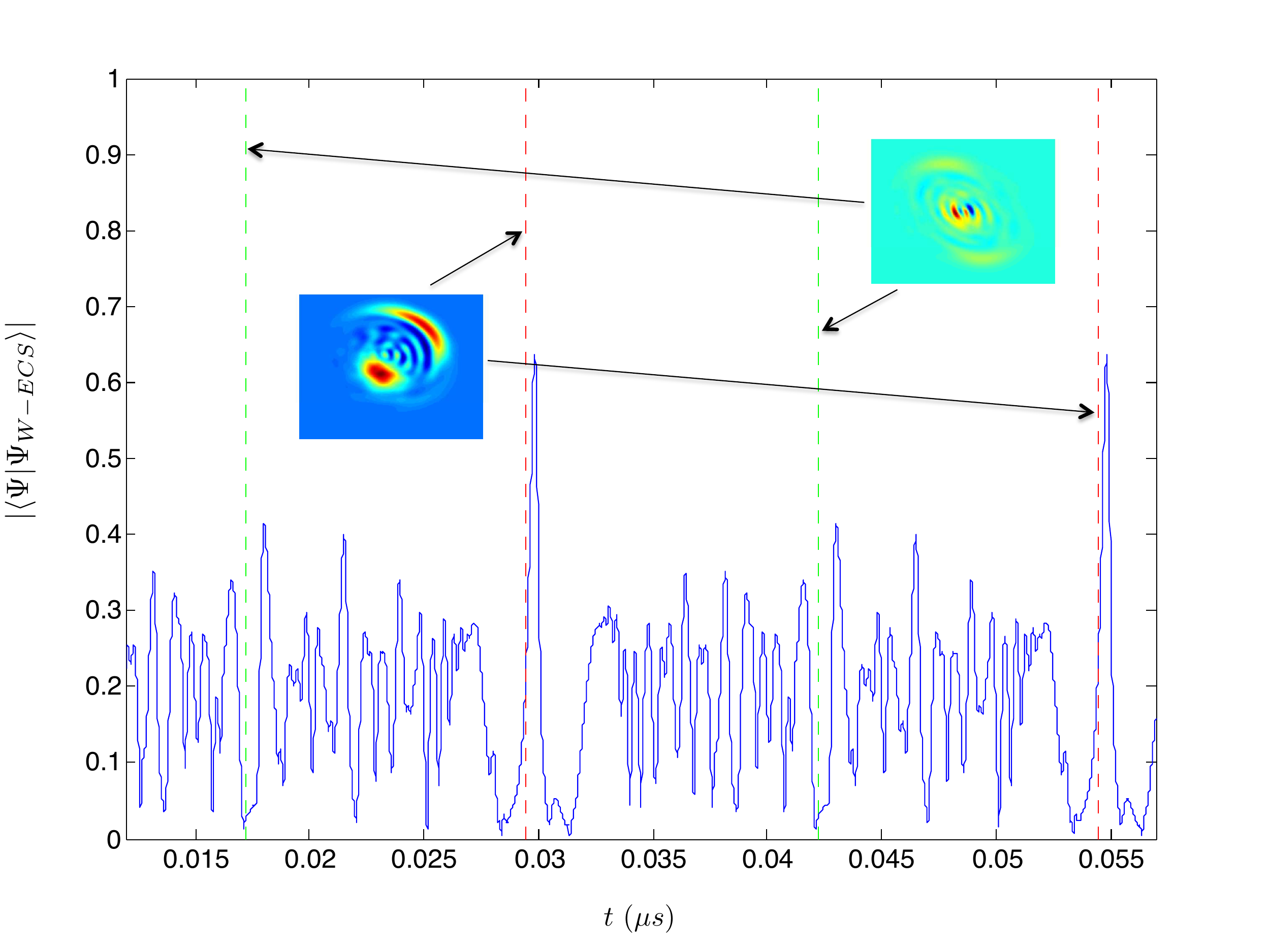}
\end{center}
\caption{Fidelity of the system state with a non-squeezed W-ECS after step two of the preparation protocol when leaving $\chi$ constant at $\chi_{max}$.  The initial state before step one is a coherent state of amplitude $|\alpha|^2=10$ in the lowest energy normal mode of a two site chain of CPWs.  The state of the system after step two periodically passes through a W-sECS (red dashed line) and W-sESCS (green dashed line) as revealed by the Wigner functions of the density matrix for the first site obtained by tracing the system density matrix over the second site at the appropriate times.}
\label{fig_step3_no_tuning} 
\end{figure}

\begin{figure} [htp]
\begin{center}
\includegraphics[scale=0.5]{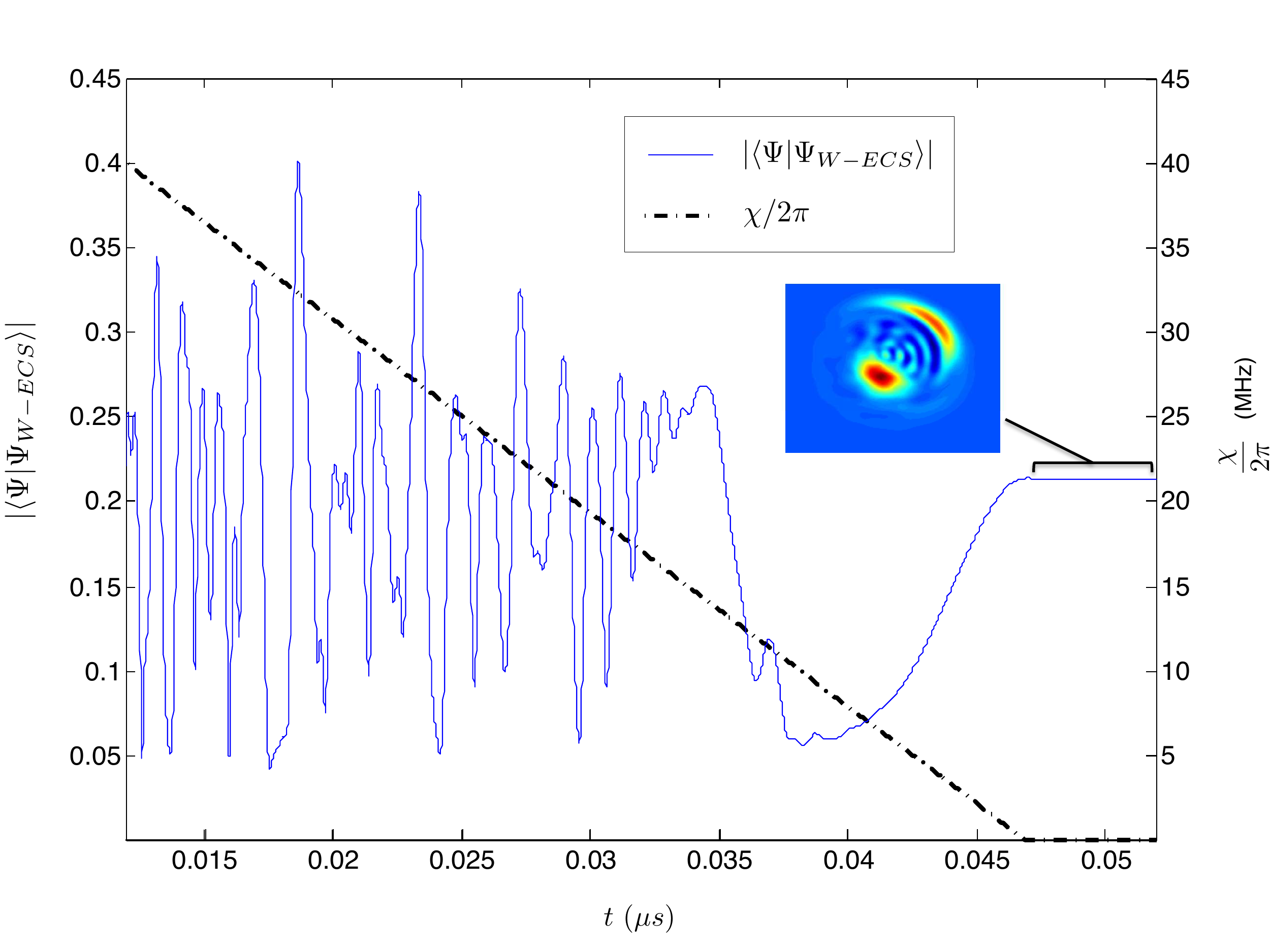}
\end{center}
\caption{Fidelity of the system state with a non-squeezed W-ECS during step three of the preparation protocol when $\chi$ is tuned from $\chi_{max}$ to zero to create the W-sECS.  The initial state before step one is a coherent state of amplitude $|\alpha|^2=10$ in the lowest energy normal mode of a two-site chain of CPWs.  The timing is chosen such that the stationary state of the system after the tuning period is a W-sECS, as indicated by the Wigner function of the first site obtained by tracing the system density matrix over the second site at the end of the tuning period.  The fidelity oscillations decrease in frequency as $\chi$ is decreased and cease when $\chi=0$.}
\label{fig_step3_ECS} 
\end{figure}

\begin{figure} [htp]
\begin{center}
\includegraphics[scale=0.5]{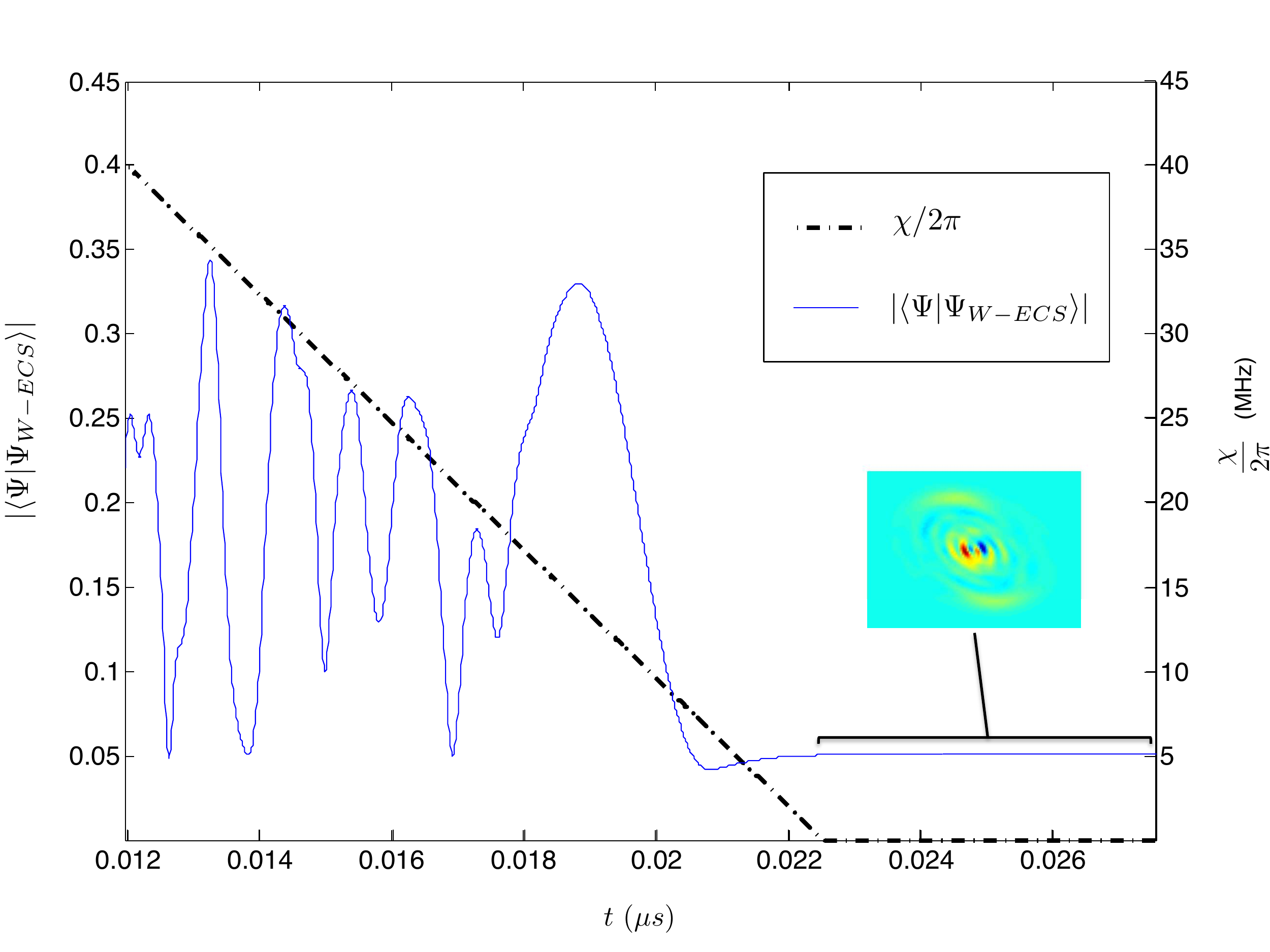}
\end{center}
\caption{Fidelity of the system state with a non-squeezed W-ECS during step three of the preparation protocol when $\chi$ is tuned from $\chi_{max}$ to zero to create the W-sESCS.  The initial state before step one is a coherent state of amplitude $|\alpha|^2=10$ in the lowest energy normal mode of a two-site chain of CPWs.  The timing is chosen such that the stationary state of the system after the tuning period is a W-sESCS, as indicated by the Wigner function of the first site obtained by tracing the system density matrix over the second site at the end of the tuning period.  The fidelity oscillations decrease in frequency as $\chi$ is decreased and cease when $\chi=0$.}
\label{fig_step3_ESCS} 
\end{figure}

Figure \ref{fig_step3_no_tuning} shows the fidelity evolution in the undamped simulation after step two if $\chi$ is left fixed at $\chi_{max}$; both the W-sECS and W-sESCS are periodically created every $2\pi/\chi=25$ ns at the indicated separate times, and their respective Wigner functions in the insets are from the density matrix of the first site obtained by tracing the system density matrix over the second site.  This indicates that by appropriately timing the tuning of $\chi$ to zero in step three, either the W-sECS or the W-sESCS can be selected as the final state of the system: $\int_0^{T^*}dt'\chi_{max}=\int_0^{T_{s3}}dt'\chi(t')$, where $T^*$ is the time at which the desired final state occurs in step three when $\chi$ is not tuned, and $T_{s3}$ is the time over which $\chi$ should be tuned from $\chi_{max}$ to zero in step three to achieve the desired final state.  For a tuning of $\chi$ linear in time we find that $T_{s3}=35$ ns for the W-sECS and $T_{s3}=10.6$ ns for the W-sESCS.  This is shown in the damped simulation results of Fig. (\ref{fig_step3_ECS}) and Fig. (\ref{fig_step3_ESCS}), respectively.  In both cases the fidelity oscillations decrease in frequency as $\chi$ is decreased, indicating a slowing of the cyclic phase shearing, until the oscillations cease when $\chi=0$.

In Supplementary Figure 2 we show that the single particle correlator $\langle c_1^{\dagger}c_2 \rangle$ generally behaves as expected for the case of W-sECS creation, but with anomalies due to imperfect adiabaticity during step one, as demonstrated in Supplementary Figure 3.  In Supplementary Figure 4, we show that  less than twenty percent of the state fidelity is lost due to damping over steps one, two, and three of the preparation protocol.

~
~
\section*{detection}

Detection of the W-sECS/W-sESCS/GHZ-sECS may be done by swapping it into the $a_j$ modes and using tunably coupled superconducting qubits (as in the case of \cite{Allman}) for bipartite Wigner tomography \cite{WangNOON} between different pairs of sites.  Alternatively, the scheme of Tuferelli et al. \cite{Tuferelli} may be used wherein the initial state of a linear oscillator network is reconstructed through readout of a single qubit attached to a single oscillator in the network.  See Supplementary Information.

\section*{Conclusions and outlook}
We have proposed a way to use existing superconducting circuit technology to experimentally simulate the dynamical manipulation of a many-body Hamiltonian for the purpose of producing a multipartite squeezed ECS of the W- or GHZ-type in the microwave domain.  Our proposal presents a new paradigm for deterministic ECS generation that is highly relevant in the face of on-going developments in superconducting circuit-based fundamental quantum tests, quantum simulation, and quantum information processing.  With appropriate timing of our protocol, the squeezed ECS may take the form of a non-locally superposed squeezed coherent state or a non-locally superposed squeezed cat state.  We have further shown how to combine existing electromechanical technology with the proposed superconducting circuit to convert the multipartite photonic quantum state into a purely mechanical one, thereby enabling the simultaneous observation of three thus far unobserved quantum phenomena in mechanical systems: multipartite non-gaussian entanglement, quadrature squeezing below the standard quantum limit, and Schr\"{o}dinger cats.  This new approach to the preparation of entangled mechanical oscillators is one that is more readily scalable to many oscillators and thereby more promising for applications.  Also, because our protocol accesses large ($|\alpha|^2\gg1$) squeezed coherent state mechanical Schr\"{o}dinger cats, it may prove useful for fundamental tests in the mechanical realm.  At a more general level, our proposal highlights the use of quantum phase transitions as a resource for fundamental and applied studies in engineered quantum systems, especially in the case of multipartite entanglement.

\section*{Methods}
The numerical simulations are performed by using the fourth order Runge-Kutta integration method on the master equation:
\begin{equation}
\frac{d\rho}{dt}=-\frac{i}{\hbar}[H,\rho]+\sum_j\frac{1}{T_1}{\cal D}[c_j]\rho+\sum_j\frac{1}{T_{\phi}}{\cal G}[c_j]\rho,
\label{master-eqn}
\end{equation}
where $\rho$ is the density matrix for the CPW chain, $H$ is the Hamiltonian of the ABH model from equation (\ref{ABH_Ham}), ${\cal D}[c_j]\rho=c_j\rho c_j^\dagger -c_j^\dagger c_j\rho/2-\rho c_j^\dagger c_j/2$ is the amplitude damping operator, ${\cal G}[c_j]\rho=c_j^\dagger c_j\rho c_j^\dagger c_j -(c_j^\dagger c_j)^2\rho/2-\rho (c_j^\dagger c_j)^2/2$ is the phase damping operator, $1/T_1$ is the amplitude damping rate, and $1/T_{\phi}$ is the phase damping rate.  The Hilbert space at each site is truncated at $n_{max}=20$, and a timestep of size $10^{-11}$ s is used.

~
~
~
~
~
~
~
~
~
~
~
~
~
~
~
~

\acknowledgments
We acknowledge the support of the Australian Research Council Centre of Excellence for Engineered Quantum Systems (grant number CE110001013).  AAG acknowledges support from the UQRS/UQIRTA scholarships and helpful discussions with Jerome Bourassa.


\begin{references}

\bibitem{Cleland} A. N. Cleland and M. L. Roukes, in Proceedings ICPS-24, edited by D. Gershoni (World Scientific, Singapore, 1999).
\bibitem{Bose} S. Bose, K. Jacobs and P. L. Knight, Phys. Rev. A 59, 3204 (1999).
\bibitem{Blencowe1} M. P. Blencowe and M. L. Wybourne, Physica B 280,555 (2000).
\bibitem{Carr} S.M. Carr, W.E. Lawrence, and M.N. Wybourne, Phys. Rev. B 64, 220101 (2001).
\bibitem{Armour} A. D. Armour, M. P. Blencowe and K. C. Schwab, Phys. Rev. Lett. 88, 148301 (2002).
\bibitem{Marshall} W. Marshall et al., Phys. Rev. Lett. 91, 130401 (2003).
\bibitem{Blencowe2} M. Blencowe, Physics Reports 395, 159 (2004).
\bibitem{Schwab} K. C. Schwab and M. L. Roukes, Phys. Today 58, 36 (2005).

\bibitem{O'Connell} A. D. O'Connell et al., Nature 464, 697 (2010).

\bibitem{Leggett} A. J. Leggett, Contemporary Physics, 25:6, 583 (1984); A. J. Leggett, J. Phys.: Condens. Matter 14, R415 (2002).
\bibitem{Mancini} S. Mancini, V. Giovannetti, D. Vitali and P. Tombesi,  Phys. Rev. Lett. 88 120401 (2002). 

\bibitem{TianCarr} L. Tian and S. M. Carr, Phys. Rev. B 74, 125314 (2006).
\bibitem{metrology} W. J. Munro, K. Nemoto, G. J. Milburn, and S. L. Braunstein, Phys. Rev. A 66, 023819 (2002).
\bibitem{BAE} V. B. Braginsky and F. Ya. Khalili, {\em Quantum Measurement}, Cambridge University Press, Cambridge, 1992; M. F. Bocko and R. Onofrio, Rev. Mod. Phys. 68, 755 (1996).
\bibitem{Zurek} W. H. Zurek, Rev. Mod. Phys., 75, 715 (2003).
\bibitem{Schlosshauer} M. Schlosshauer, A. P. Hines, G. J. Milburn, Phys. Rev. A 77, 022111 (2008).
\bibitem{Remus} L. G. Remus, M. P. Blencowe, Y. Tanaka, Phys. Rev. B 80, 174103 (2009).

\bibitem{Bassi} A. Bassi et al., arXiv:1204.4325.
\bibitem{Pepper} B. Pepper et al., arXiv:1207.1946.
\bibitem{Sanders_review} B. C. Sanders, J. Phys. A: Math. Theor. 45, 244002 (2012).
\bibitem{SandersECS} B. C. Sanders, Phys. Rev. A 45, 6811-6815 (1992); B. C. Sanders, Phys. Rev. A 46, 2966 (1992).

\bibitem{BoseAgarwal} S. Bose and G. S. Agarwal, New Journal of Physics 8, 34 (2006).
\bibitem{TianZoller} L. Tian and P. Zoller, Phys. Rev. Lett. 93, 266403 (2004).
\bibitem{ECSvsEPP} W. J. Munro, G. J. Milburn, and B. C. Sanders, Phys. Rev. A 62, 052108 (2000); K. Park and H. Jeong, Phys. Rev. A 82, 062325 (2010).
\bibitem{opticalECS} A. Ourjoumtsev, Nature Physics 5, 189 (2009).
\bibitem{W-state decay} R. Chaves and L. Davidovich, Phys. Rev. A 82, 052308 (2010); A. Montakhab and A. Asadian, Phys. Rev. A 77, 062322 (2008); A. R. R. Carvalho, F. Mintert, and A. Buchleitner, Phys. Rev. Lett. 93, 230501 (2004).
\bibitem{Braunstein} Jing Zhang, Kunchi Peng, and Samuel L. Braunstein, Phys. Rev. A 68, 013808 (2003).

\bibitem{BECcat} J. I. Cirac, M. Lewenstein, K. Molmer, and P. Zoller, Phys. Rev. A 57, 1208 (1998).
\bibitem{Lee} C. Lee, Phys. Rev. Lett. 97, 150402 (2006).
\bibitem{Plenio} M. J. Hartmann, F. G. S. L. Brandao, and M. B. Plenio, Nat. Phys. 2, 849 (2006).

\bibitem{Houck} A. A. Houck, H. E. Tureci, and J. Koch, Nature Physics 8, 292 (2012).
\bibitem{MECSnote} This is in contrast to the scheme of Wang and Sanders \cite{MECS}, which considers how to generate an ECS amongst the different normal modes of a single chain of coupled ions.
\bibitem{MECS} X. Wang and B. C. Sanders, Phys. Rev. A 65, 012303 (2001).
\bibitem{Teufel1} J. D. Teufel et al., Nature {\bf 471}, 204 (2011).
\bibitem{Teufel2} J. D. Teufel et al., Nature {\bf 475}, 359 (2011).
\bibitem{transmon_expts} J. M. Fink et al., Nature {\bf 454}, 315 (2008).  
\bibitem{Bourassa} J. Bourassa, F. Beaudoin, J. M. Gambetta, and A. Blais, Phys. Rev. A 86, 013814 (2012).
\bibitem{Palomaki} T. A. Palomaki et al., arXiv:1206.5562.
\bibitem{coupler} R. C. Bialzcak et. al, Phys. Rev. Lett. 106, 060501 (2011).
\bibitem{coupler2} Y. Yin et al., arXiv:1208.2950.
\bibitem{WangNOON} H. Wang et al., Phys. Rev. Lett. 106, 060401 (2011).
\bibitem{transmon1} A. A. Houck et al., Quantum Inf. Process. 8, 105 (2005).
\bibitem{Bourassa2} J. Bourassa, private communication.

\bibitem{attractiveBH1} M. W. Jack and M. Yamashita, Phys. Rev. A 71, 023610 (2005).
\bibitem{attractiveBH2} P. Buonsante et al., Phys. Rev. A 72, 043620 (2005).
\bibitem{attractiveBH3} P. Buonsante et al.,  J. Phys. B (2006).
\bibitem{attractiveBH4} N. Oelkers and J. Links, Phys. Rev. B 75, 115119 (2007).
\bibitem{attractiveBH5} P. Buonsante et al., Phys. Rev. A 82, 043615 (2010).
\bibitem{attractiveBH6} J. I. Cirac, Phys. Rev. A 77, 033403 (2008).
\bibitem{attractiveBH7} G. Mazzarella, L. Salasnich, A. Parola, and F. Toigo, Phys. Rev. A 83, 053607 (2011).
\bibitem{attractiveBH8} M. J. Steel and M. J. Collett, Phys. Rev. A 57, 2920 (1998).
\bibitem{attractiveBH9} P. Zin, Euro. Phys. Lett. 83, 64007 (2008).
\bibitem{Lieb} M. Lieb and M. J. Hartmann, New J. Phys. 12, 093031 (2010).
\bibitem{BL} S. Lloyd and S. L. Braunstein, Phys. Rev. Lett. 82, 1784 (1999).
\bibitem{QuantumOptics} D. F. Walls and G. J. Milburn, {\em Quantum Optics} (1994).
\bibitem{Kerr_experiment} G. Kirchmair et al., arXiv:1211:2228.

\bibitem{WECS1} N. B. An, Phys. Rev. A 69, 022315 (2004).
\bibitem{WECS2} H. Jeong and N. B. An, Phys. Rev. A 74, 022104 (2006).
\bibitem{WECS3} M.-F. Chen and S.-S. Ma, Acta Photon. Sin. 36950 (2007).
\bibitem{WECS4} Y. Guo and L. M. Kuang, J. Phys. B: At. Mol. Opt. Phys. 40, 3309 (2007).
\bibitem{hybridmode} A third dynamical process is also present as the modes $c_j$ are actually hybrid SQUID-cavity modes, but the SQUID-cavity coupling of $\sim0.1~\omega_c$ \cite{Bourassa} is large enough so that the tuning of $\omega_c$ that is concomitant with the tuning of $\chi$ can always be considered adiabatic with respect to the internal dynamics of the $c_j$ modes.
\bibitem{Hofheinz} M. Hofheinz et al., Nature {\bf 454}, 310 (2008).
\bibitem{Underwood} D. Underwood et al., Phys. Rev. A 86, 023837 (2012).

\bibitem{Allman} M. S. Allman, F. Altomare, J. D. Whittaker, K. Cicak, D. Li, A. Sirois, J. Strong, J. D. Teufel, and R.W. Simmonds, Phys. Rev. Lett. 104, 177004 (2010).
\bibitem{Tuferelli}  T. Tufarelli et al., Phys. Rev. A 85, 032334 (2012).



\end{references}
\end{document}


\title{Supplementary Information for "Non-locally entangled microwave and micromechanical squeezed cats: a phase transition-based protocol"}

\author{A. A. Gangat}
\affiliation{ARC Centre for Engineered Quantum Systems, School of Mathematics and Physics, The University of Queensland, St Lucia, QLD 4072, Australia}
\author{I. P. McCulloch}
\affiliation{ARC Centre for Engineered Quantum Systems, School of Mathematics and Physics, The University of Queensland, St Lucia, QLD 4072, Australia}
\author{G. J. Milburn}
\affiliation{ARC Centre for Engineered Quantum Systems, School of Mathematics and Physics, The University of Queensland, St Lucia, QLD 4072, Australia}

\maketitle

\section{CONSTRAINT ON COHERENT STATE AMPLITUDE TO NEGLECT HIGHER ORDER TERMS OF CPW NONLINEARITY}

The full expression for the nonlinearity ($H_{NL}$) introduced into the CPWs by the transmon qubits is given in equation 27 of \cite{Bourassa}.  Looking only at the fundamental mode, and using the relations $\Phi_0=h/2e$, $E_J=(\frac{\Phi_0}{2\pi})^2/L_J$, $L_J=L'/\eta_{l}$, and $L'=2E'_{C}/(\omega_{c,0}^2e^2)$ we find
\begin{equation}
H_{NL} = \sum_{i>1} \frac{(-1)^{i+1}}{2(2i!)}\Big(\frac{4\chi}{\omega_{c,0}}\Big)^{i-1}\eta_l~\hbar\omega_{c,0}[c^{\dagger}+c]^{2i},
\end{equation}
where $\eta_l$ ($0\lesssim\eta_l\lesssim1$) is the inductive participation ratio dependent upon the flux through the SQUID loop and $\omega_{c,0}$ is the fundamental mode frequency in the absence of the transmon.  After the rotating wave approximation, to be able to neglect nonlinear terms higher than $(c^{\dagger}c)^2$ we require 
\begin{equation}
\frac{1}{12}\Big(\frac{\chi}{\omega_{c,0}}\Big)6\Big(1.5\langle c^{\dagger}c\rangle\Big)^{2}\gg\frac{1}{30}\Big(\frac{\chi}{\omega_{c,0}}\Big)^2 20\Big(1.5\langle c^{\dagger}c\rangle\Big)^{3},
\end{equation}
or
\begin{equation}
\chi\ll\frac{1}{2}\frac{\omega_{c,0}}{|\alpha|^2},
\end{equation}
where we use the fact that greater than 95$\%$ of a coherent state with amplitude $\alpha$ occupies number states with $n\leq1.5|\alpha|^2$ for $|\alpha|^2\geq10$.  Using $\chi=40$ MHz, $\omega_{c,0}=7.5$ GHz, and $|\alpha|^2=10$, we find that the constraint is satisfied.  We note that larger values of $\chi$ and/or $|\alpha|^2$ can be accommodated by increasing the bare CPW fundamental mode frequency $\omega_{c,0}$ as long as the lumped LC oscillator frequency $\omega_a$ matches $\omega_{c,0}$.  Larger values of $\chi$ in turn allow for larger values of $\kappa''_{max}$ and therefore smaller $\Delta t_1$ and $\Delta t_3$, thereby increasing the quantum coherence of the state when transferred to the mechanical modes.

\section{BOSONIC SIMULATOR DEMONSTRATION FOR BIPARTITE W-ECS CREATION}
Consider for the case of two sites the Hamiltonian given in equation (5) of the main article, 
\begin{equation}
\frac{H}{\hbar} = -\frac{\chi_1(t)}{2}c_1^{\dagger}c_1(c_{1}^{\dagger} c_{1} - 1) -\frac{\chi_2(t)}{2}c_2^{\dagger}c_2(c_{2}^{\dagger} c_{2} - 1) - \kappa''(t)(c_{1}^{\dagger} c_{2} +  c_{1}c_{2}^{\dagger}),
\end{equation}
with the explicit assumption that each of the quartic non-linear terms and the quadratic coupling term may be subject to independent control. 

We begin by assuming that the Kerr nonlinearity is set to zero. Then both normal modes, $A_1=(a_1+a_2)/\sqrt{2},\ \ \ A_2=(a_1-a_2)/\sqrt{2}$,  of the coupled cavity system can each be coherently driven to a steady-state coherent state. The state of the local modes is likewise a product coherent state, and by a suitable choice of the normal mode driving fields it becomes $|\alpha\rangle_{a_1}|-\alpha\rangle_{a_2}$. In the first step we set $\kappa''$ to  zero, to decouple the cavities,  and the Kerr nonlinear term for cavity $a_1$  is to a non-zero value for a fixed time $T_1$ such that $\chi T_1=\pi$. The Kerr nonlinearity for mode $a_2$ remains set at zero.  This transforms the local modes as 
\begin{equation}
|\alpha\rangle_{a_1}|-\alpha\rangle_{a_2}  \rightarrow \frac{1}{\sqrt{2}}\left (e^{-i\pi/4}|i\alpha\rangle_{a_1}+e^{i\pi/4}|-i\alpha\rangle_{a_1}\right )|-\alpha\rangle_{a_2}
\end{equation}
In the next step we turn on the coupling between the cavities for a fixed amount of time $T_2$ so that $\kappa'' T_2=\pi/4$, the resulting state is 
\begin{equation}
e^{-i\pi/4}|i\sqrt{2}\alpha\rangle_{a_1}|0\rangle_{a_2}+e^{i\pi/4}|0\rangle_{a_1}|-\sqrt{2}\alpha\rangle_{a_2}
\end{equation}
While this is not precisely a W-ECS it is a non-classical entangled cat state. Using suitable phase rotations and displacements  however it can easily be transformed into a W-ECS state.

\section{REMAINING COHERENCE FRACTION AFTER STATE PREPARATION}

\begin{figure} [htp]
\begin{center}
\includegraphics[scale=.8]{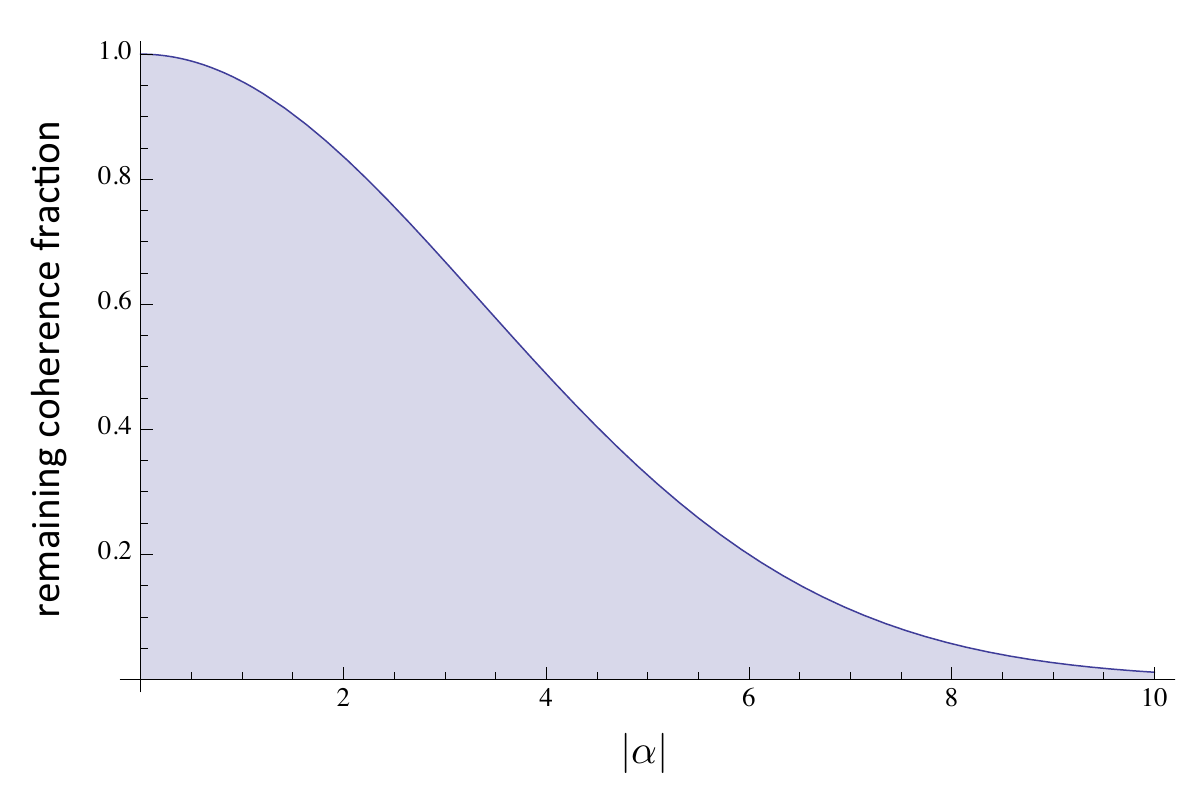}
\end{center}
\caption{Approximate remaining fraction of quantum coherence as a function of normal mode coherent state amplitude $|\alpha|$ of initial M-partite W-ECS ($M\lesssim10$) after transfer to mechanical modes.  The calculation is done under the approximation that the coherent state is already in ECS form at the beginning of step one.}
\label{figure_coherence} 
\end{figure}

We can estimate the quantum coherence of the ECS after it is prepared and transferred to the mechanical modes.  Using the formula $\frac{1}{T_{0n}}=\frac{n}{T_{1}}+\frac{n^2}{T_{\phi}}$ (see \cite{dampedNOON}), an average $T_1\approx2~\mu$s, and an average dephasing time of $T_{\phi}\approx0.1$ s over the relevant flux range, we estimate the average decoherence over steps one, two, three, and four for a coherent superposition $|n\rangle|0\rangle + |0\rangle|n\rangle$ across two of the $c_j$ modes to be $D_{1,2,3,4}(n)=e^{-(\Delta t_1+\Delta t_2+\Delta t_3+\Delta t_4)(n/(2\times10^{-6})+n^2/(100\times10^{-6})}$.  The decoherence $D_{5a}(n)=e^{-\Delta t_{5a}/T_{0n}}$ over $\Delta t_{5a}$ entails an average of the decay rates of the CPW and the LC oscillator (we assume dephasing to be negligible for both modes when $\chi=0$): $T_{0n}=2/n(\gamma_c+\gamma_a)~s=3.42/n~\mu s$.  The decoherence $D_{5b}(n)=e^{-\Delta t_{5b}/T_{0n}}$ over $\Delta t_{5b}$ entails an average of the decay rates of the LC oscillator and the mechanical mode (we assume dephasing to be negligible for these modes): $T_{0n}=2/n(n_{env}\gamma_b+\gamma_a)~s=7.67/n~\mu s$, where $n_{env}=50$ is due to the mechanical mode thermal environment at 25 mK \cite{Palomaki}.  In Fig. (\ref{figure_coherence}) we plot the total remaining quantum coherence of a bipartite W-ECS formed in a two site lattice with coherent state amplitude $\alpha$ after all of the preparation stages under the approximation that the coherent state is already in ECS form at the beginning of step one: $\sum_{n=0}^{1000} |\langle n|\alpha\rangle|^2 D_{1,2,3,4}(n)D_{5a}(n)D_{5b}(n)$  vs. $|\alpha|$.  We use $\Delta t_1=.02~\mu$s, $\Delta t_2=.002~\mu$s, $\Delta t_3=.02~\mu$s, $\Delta t_4=.002~\mu$s, $\Delta t_{5a}=.01~\mu$s, and $\Delta t_{5b}=.16~\mu$s.  The values of $\Delta t_1$ and $\Delta t_2$ accommodate $M\lesssim10$.  The plot indicates that our proposed system and protocol can achieve at each site a purely mechanical cat state with greater than sixty percent quantum coherence and an average centre of mass spatial separation of a few times the mechanical zero point motion.

\section{Behaviour of $\langle c_1^{\dagger}c_2 \rangle$ for the case of W-sECS creation}

\begin{figure} [htp]
\begin{center}
\includegraphics[scale=0.5]{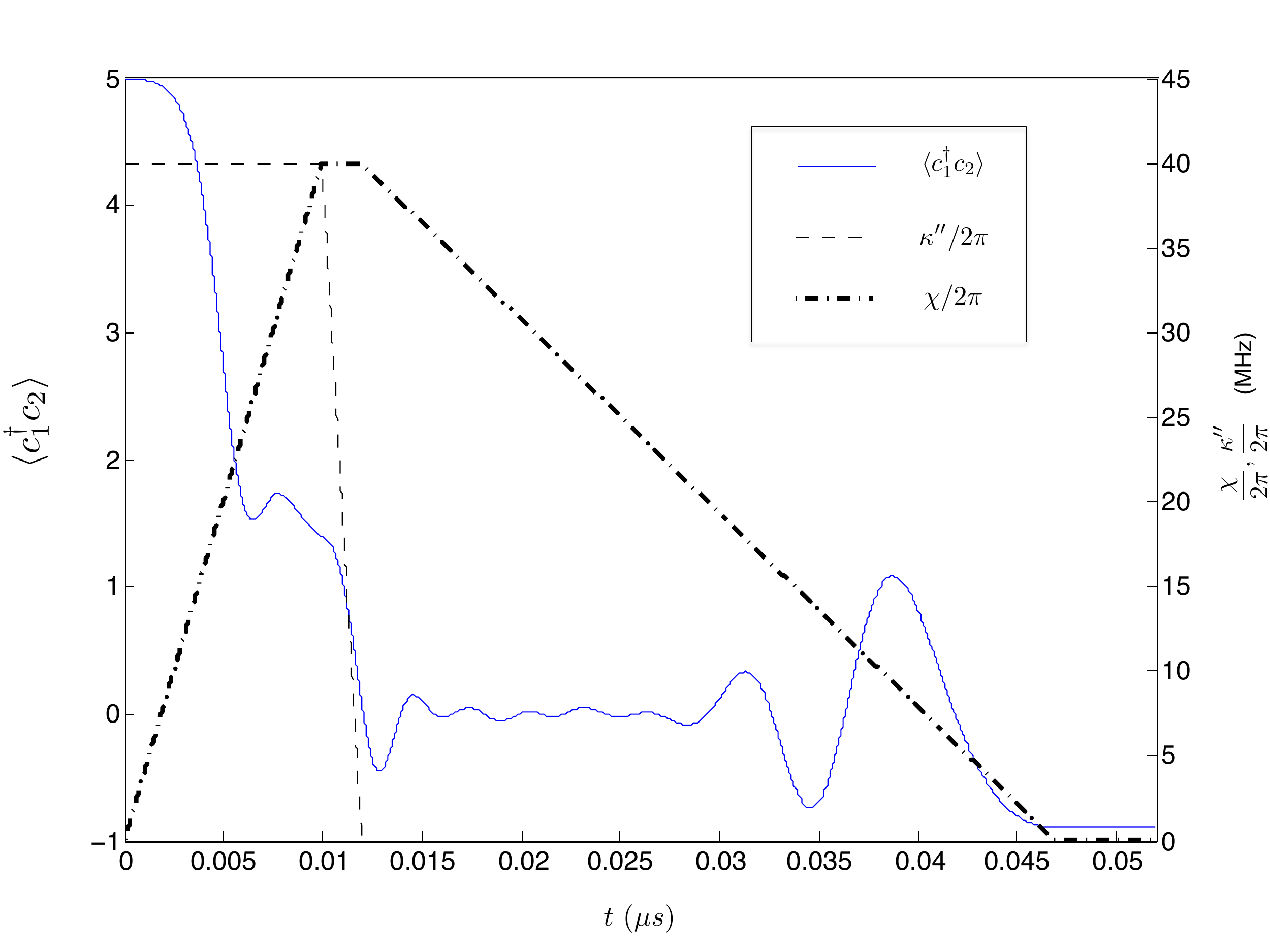}
\end{center}
\caption{$\langle c_1^{\dagger}c_2 \rangle$ over steps one, two, three, and a 5 ns period after step 3.  The oscillation pulses are discussed below.  The ideal ECS has a value of zero, indicating complete site-localization.}
\label{fig_C1} 
\end{figure}

The single particle correlator $\langle c_1^{\dagger}c_2 \rangle$ between sites one and two is plotted in Fig. \ref{fig_C1} over all three steps for the case of W-sECS creation plus an additional 5 ns free evolution period after step three.  The steady overall decrease of $\langle c_1^{\dagger}c_2 \rangle$ indicates increasing site-localization of the initial coherent state (the perfect ECS has complete site-localization and a single particle correlator value of zero).  The intermittent oscillations appear to have a frequency proportional to $\chi$, which indicates that they are a signature of dynamics between the different eigenmodes of the lattice.  This arises because the nonlinear term of the Hamiltonian induces scattering between the normal modes (as easily seen by Fourier transforming the local mode operators $c_j$), and because the tuning of $\chi$ in step one is not perfectly adiabatic.  This is verified below.  The W-type distribution of the coherent states is not affected by this because all of the low energy states in the regime $\tau\leq0.25$ are of the W-state form with different local phase variations only \cite{attractiveBH5}.

\begin{figure} [htp]
\begin{center}
\includegraphics[scale=1]{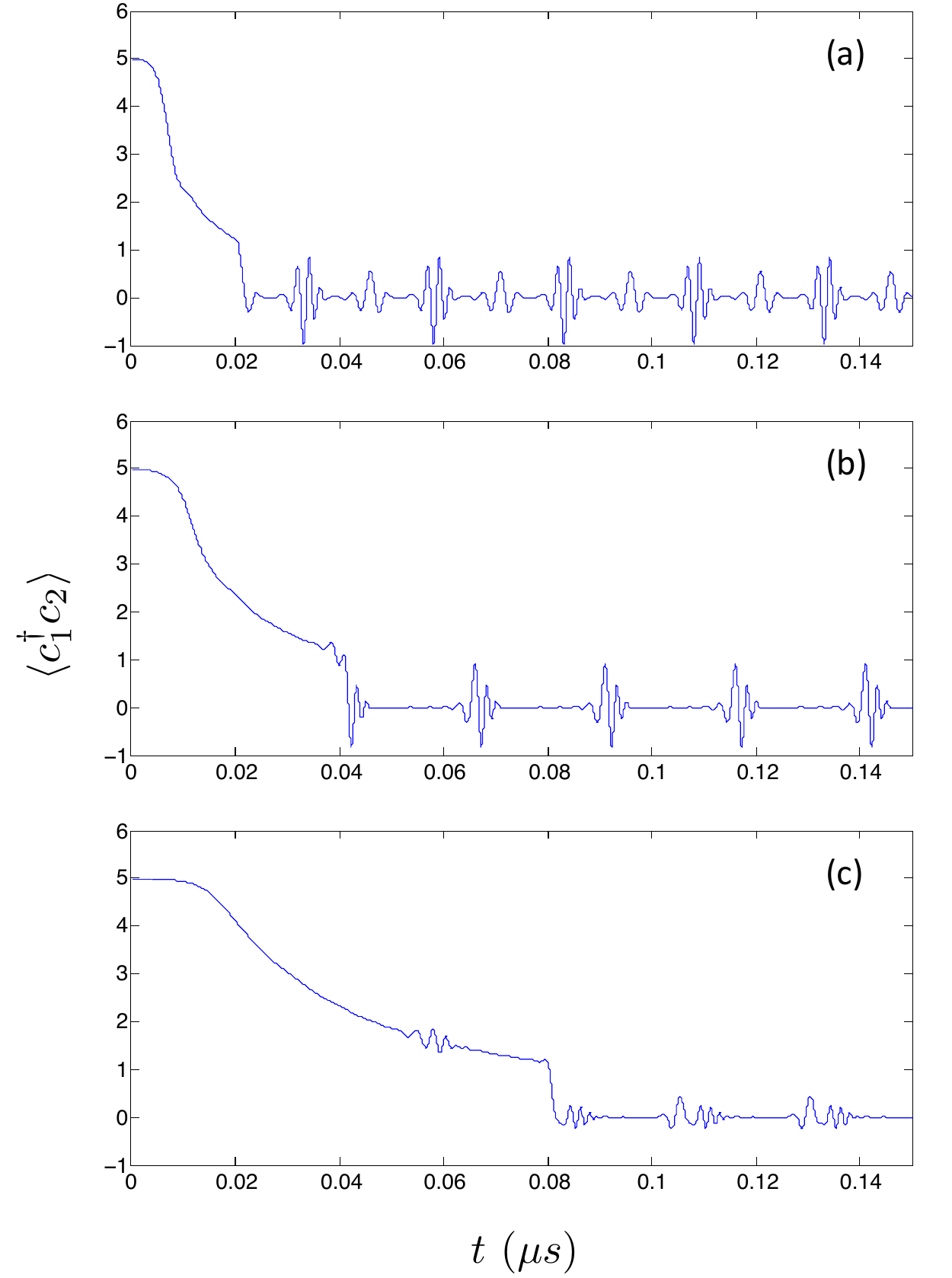}
\end{center}
\caption{Undamped simulation of $\langle c_1^{\dagger}c_2 \rangle$ over steps one and two, followed by a period of unitary evolution with $\chi=\chi_{max}$. {\bf a}, $\Delta t_1=0.02~\mu$s.  {\bf b}, $\Delta t_1=0.04~\mu$s.  {\bf c}, $\Delta t_1=0.08~\mu$s.  With increasing $\Delta t_1$ the adiabaticity of step one is improved and the oscillations are consequently increasingly damped.  The steep drop after $\Delta t_1$ is due to step two, where the sites are decoupled by tuning $\kappa''$ to zero.  Other than $\Delta t_1$, the simulation parameters are the same as for the undamped simulation of Fig. 5 in the main text.}
\label{fig_C1_adiabtest} 
\end{figure}

To show that the amplitude of the intermittent oscillations decreases as $\Delta t_1$ is increased, we perform three undamped numerical simulations of steps one and two of the state preparation protocol followed by a period of unitary evolution with $\chi=\chi_{max}$.  Each simulation uses different values of $\Delta t_1$ in order to test the dependence of the oscillations of $\langle c_1^{\dagger}c_2 \rangle$ on the adiabaticity of step one (where $\chi$ is tuned from zero to $\chi_{max}$) and the results are shown in Fig. \ref{fig_C1_adiabtest}.  The oscillation amplitudes clearly decay with increasing $\Delta t_1$, indicating that the oscillations are due to the imperfect adiabaticity of step one.

\section{Effect of amplitude and phase damping on the fidelity of the created state}

\begin{figure} [htp]
\begin{center}
\includegraphics[scale=0.5]{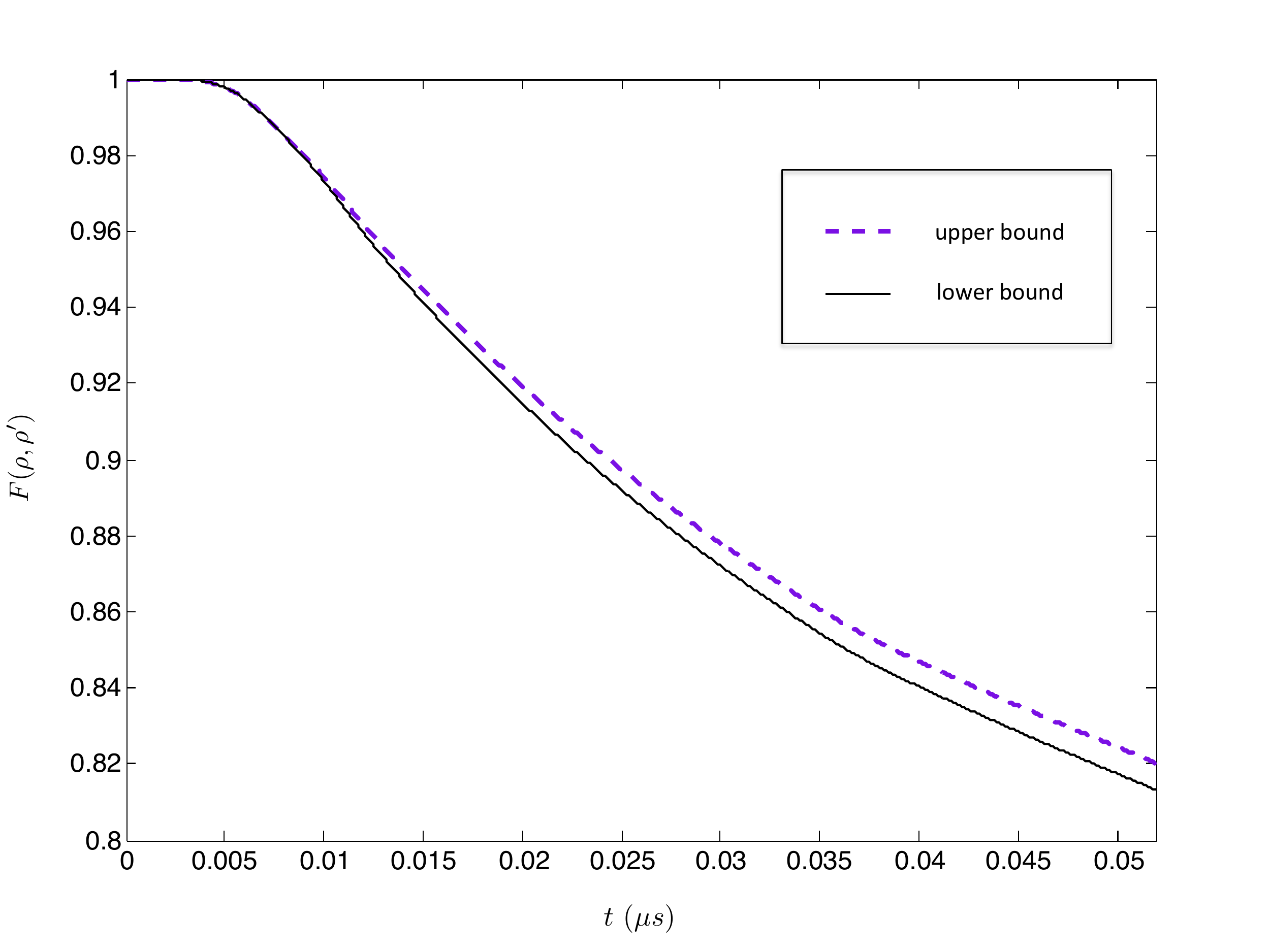}
\end{center}
\caption{Upper bound (dashed purple) and lower bound (solid black) of the fidelity of the system state $\rho$ with the undamped reference system state $\rho'$ over the preparation of the W-sECS.}
\label{fig_fidelity_damped} 
\end{figure}

To investigate the effect of the amplitude and phase damping on the state preparation, we consider the fidelity $F(\rho,\rho')=[tr\sqrt{\sqrt{\rho}\rho'\sqrt{\rho}}]^2$ of the damped system state ($\rho$) with the undamped system state ($\rho'$) through all steps of the preparation protocol for the W-sECS.  As $F(\rho,\rho')$ is computationally very expensive to compute, we calculate instead the upper bound, given by $F(\rho,\rho')\leq tr\rho\rho'+\sqrt{(1-tr\rho^2)(1-tr\rho'^2)}$, and the lower bound, given by $F(\rho,\rho')\geq tr\rho\rho'+\sqrt{2}\sqrt{(tr\rho\rho')^2-tr\rho^2\rho'^2}$ \cite{superfidelity}.  Both are shown in Fig. \ref{fig_fidelity_damped}, which clearly indicates that less than twenty percent of the fidelity is lost due to damping over steps one, two, and three of the preparation protocol.

\section{State detection via a single qubit}
After creation of the mechanical/microwave W-sECS/W-sESCS/GHZ-sECS in our system, all couplings $\kappa'$, $\kappa''$, and $g$ could be turned on while leaving $\chi=0$ to make the network linear.  In this case, the scheme of Tuferelli et al. \cite{Tuferelli} would allow for a single qubit tunably coupled to any microwave mode in our system to reconstruct the initial mechanical/microwave W-sECS/W-sESCS/GHZ-sECS of the system.  One caveat of this approach is that it requires Markovian master equation modelling of the dissipative environment for all of the oscillators of the network.  Because the correct such model for the mechanical dissipation is unknown, the bipartite Wigner tomography method may need to be used to first verify which dissipative model to use before applying the method of Tuferelli et al. to read out the entire network.